\documentclass[%
twocolumn,
 amsmath,amssymb,
aps,
prx,
longbibliography]{revtex4-1}

\usepackage{graphicx,epsfig,epstopdf}
\usepackage{amsmath}
\usepackage{color}
\usepackage{caption}
\usepackage{subcaption}
\usepackage{ulem}

\def\e{\begin{equation}}
\def\f{\end{equation}}
\def\_#1{{\bf #1}}
\def\.{\cdot}

\def\l#1{\label{eq:#1}}
\def\r#1{(\ref{eq:#1})}

\begin{document}

\title{Illusion Mechanisms with Cylindrical Metasurfaces: A General Synthesis Approach}

\author{Mahdi Safari$^{1}$, Hamidreza Kazemi$^{2}$, Ali Abdolali$^{3}$, Mohammad Albooyeh$^{2,\ast}$, and Filippo Capolino$^{2}$}
 
\address{$^1$Department of Electrical and Computer Engineering, University of Toronto, Toronto, Canada\\
$^2$Department of Electrical Engineering and Computer Science, University of California, Irvine, CA 92617, USA\\
$^3$Department of Electrical Engineering, Iran University of Science and Technology, Narmak, Tehran, Iran\\
{\rm corresponding author: $^{\ast}$mohammad.albooyeh@gmail.com}}

\begin{abstract}

We explore the use of cylindrical metasurfaces in providing several illusion mechanisms including scattering cancellation and creating fictitious line sources. We present the general synthesis approach that leads to such phenomena by modeling the metasurface with effective polarizability tensors and by applying boundary conditions to connect the tangential components of the desired fields to the required surface polarization current densities that generate such fields. We then use these required surface polarizations to obtain the effective polarizabilities for the synthesis of the metasurface. We demonstrate the use of this general method for the synthesis of metasurfaces that lead to scattering cancellation and illusion effects, and discuss  practical scenarios by using loaded dipole antennas to realize the discretized polarization current densities. This study is the first fundamental step that may lead to interesting electromagnetic applications, like stealth technology, antenna synthesis, wireless power transfer, sensors, cylindrical absorbers, etc.

\end{abstract}

\maketitle

\section{Introduction}

Metasurfaces are surface equivalents of bulk metamaterials, usually realized as dense planar arrays of subwavelength-sized resonant particles~\cite{Kuester, Holloway1, Holloway2, explain2}. Since the emergence of this research topic, most studies have been carried out to analyze and synthesize infinitely extended planar metasurfaces~\cite{kazemi2019simultaneous,Padilla, Caiazzo, Chan,Pfeiffer1,Selv,Capolino,Pfeiffer2, Zhao,Albooyeh1,Albooyeh2,Asadchy1,
Asadchy,Epstein1,Epstein2,Roberts, Dana,Albooyeh2018Applications,Niemi,Salem,Kazemi2019Perfect,Achouri,Albooyeh3,Albooyeh31,
Kiani,Oraizi,Alaeeabs}. The synthesis of some topologies other than planar, namely cylindrical and spherical ones, has been studied in the past~\cite{Kishk,Raffaelli,Padooru,Padooru1,Bernety,jia,Dehmol,
Raeker,Raeker2} but without a general systematic approach. Cylindrical topologies are among the commonly used structures in engineering electromagnetics and optics. For instance, cloaking, radar cross section reduction, obtaining an arbitrary radiation pattern from cylindrical objects, and etc. are only a few problems which involve metasurfaces synthesis in cylindrical coordinates.\\
For the analysis and synthesis of planar metasurfaces, several techniques have been reported as in Refs~\cite{Pfeiffer1,Selv,Capolino,Pfeiffer2,
Zhao,Albooyeh1,Albooyeh2,Asadchy1,
Asadchy,Epstein1,Epstein2,Dana,
Niemi,Salem,Achouri,Albooyeh3,Albooyeh31,Roberts} for example. These works are based on modeling the metasurfaces with impedance tensors~\cite{Pfeiffer1,Selv,Capolino,Pfeiffer2,
Zhao,Albooyeh1,Albooyeh2,Asadchy1,
Asadchy,Epstein1,Epstein2}, with equivalent conductivities and reactances~\cite{Dana}, or with effective surface susceptibility or polarizability  tensors~\cite{Niemi,Salem,Achouri,Albooyeh3,Albooyeh31,
Roberts}. Although the analysis and synthesis of planar metasurfaces are well-established and discussed in the literature, to the best of our knowledge, such a comprehensive analysis and/or synthesis is not yet conducted for cylindrical metasurfaces. Therefore, we believe there is a need to develop an extensive analysis and synthesis method for assigning prescribed physical properties to cylindrical metasurfaces.\\
In particular here we aim at developing some illusion mechanisms using cylindrical metasurfaces. To that end, we present a comprehensive analysis and synthesis method for cylindrical metasurfaces analogous to what has been previously performed for the case of planar metasurfaces in Refs.~\cite{Albooyeh3,Albooyeh31}. Note that the analysis of planar metasurfaces has been already extended to conformal metaurfaces with large radial curvatures (at the wavelength scale)~\cite{Tretyakov}. However, that method was based on the analysis of planar structures with open boundaries while a cylindrical metasurface can be generally closed in its azimuthal plane and it involves rather different interaction mechanisms. Our formulation covers the most general case of metasurfaces with dipolar (electric and magnetic) resonant responses including bianisotropic cases~\cite{Serdyukov}. By applying our synthesis approach, we further conceptually synthesize several examples with interesting practical applications, e.g., scattering cancellation from conductive and dielectric cylinders and also generating fictitious line sources, appearing away from the original line source.\\
The paper is structured as follows. We first present the theory for the analysis of infinitely extended (along the axial direction) cylindrical metasurfaces in Secs.~\ref{PF} by providing the relation between the desired fields at the metasurface boundary and the required surface polarization densities. Next, we exhibit an expansion of the incident (here, plane waves and line sources) and scattered fields in cylindrical coordinates in Secs.~\ref{incindentfield} and~\ref{scattfields}. After that, we conceptually synthesize two examples of scattering cancellation and an illusion device in Secs.~\ref{Excloak},~\ref{Excloak2}, and~\ref{Exillusion}. We further present a practical realization approach by discretizing the continuous required polarization currents in Sec.~\ref{PR}. Finally, we conclude the study by discussing the influence of the presented method in practical applications and new possibilities by cylindrical metasurfaces in engineering electromagnetic waves.

\section{Problem formulation and boundary conditions}\label{PF}

\begin{figure}[h!]
\centering
 \epsfig{file=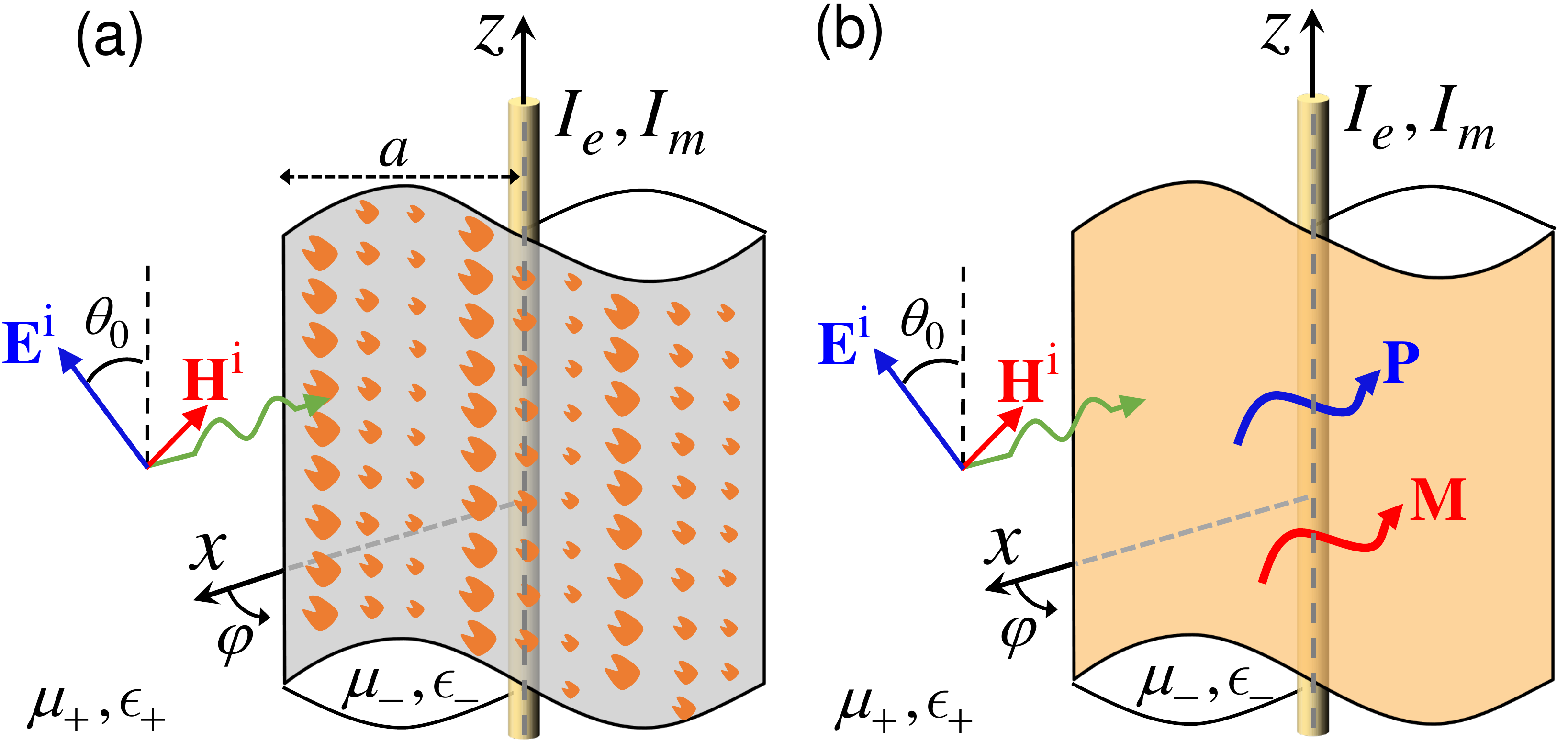, width=0.99\linewidth}  
  \caption{(a) A cylindrical metasurface composed of nonidentical inclusions which is infinitly extended in the axial direction (here $z$ axis) and is exposed to incidences from inside (i.e., infinite line sources with electric and magnetic current amplitudes $I_e$ and $I_m$) and outside (i.e., a plane wave). (b) The electromagnetic modeling of the problem in (a). The metasurface is replaced by two surface polarization densities $\_P$ and $\_M$ that provide the same electromagnetic response as the problem in (a).}\label{geom}
\end{figure}

An electromagnetic metasurface is commonly understood as a composite layer composed of infinite number of subwavelength-sized inclusions which are densely arranged over a surface and is capable of manipulating the wave front in a desired fashion (see e.g. Refs.~\cite{Holloway2}). This definition includes also the case of non-planar conformal surfaces. Here, we study a class of metasurfaces when the inclusions are arranged to form a cylindrical surface which is closed in the azimuthal plane and is infinitely extended in the axial direction. Note that in the limiting case when the cylinder curvature is infinitely large, the current problem transitions to the case of planar metasurfaces which is widely studied (see e.g. Refs.~\cite{Albooyeh2,Roberts,Dana,Niemi,Salem,Kazemi2019Perfect,Achouri,
Albooyeh3,Albooyeh31}). However, in sharp contrast with planar metasurfaces, in a cylindrical configuration waves bounce inside the cylindrical metasurface and multiple reflections form the inner part of the cylindrical boundaries must be considered in the analysis.\\
Let us consider a metasurface composed of resonant inclusions that form a cylindrical surface with radius $a$ [see Fig.~\ref{geom}].
The cylindrical surface is assumed to be infinitely extended along the axial direction (here $z$-direction). Moreover, we assume that the inner and outer spaces of the cylindrical surface are filled with two different isotropic materials with the permittivity $\epsilon_{\pm}$ and permeability $\mu_{\pm}$, respectively (see Fig.~\ref{geom}, the ``$+$'' sign refers to the material properties of the outside medium whereas the ``$-$'' sign corresponds to the inside one). We illuminate the proposed metasurface either from outside or from inside. Due to the external/internal illumination, the effective surface electric and magnetic polarization densities ${\bf P}$ (with the unit of $\rm Asm^{-1}$) and ${\bf M}$ (with the unit of $\rm Vsm^{-1}$) are, respectively, induced on the metasurface boundary (see e.g.~\cite{Niemi,Salem,Kazemi2019Perfect,Achouri,Albooyeh3,Albooyeh31}). Next, similar to Refs.~\cite{Niemi,Salem,Achouri,Albooyeh3,Albooyeh31}, we express the metasurface response in terms of  {\it effective} electric, magnetic, magnetoelectric, and electromagnetic polarizability tensors ${\bar{\bar{\alpha}}}^\textrm{ee}$, ${\bar{\bar{\alpha}}}^\textrm{mm}$,${\bar{\bar{\alpha}}}^\textrm{em}$, and ${\bar{\bar{\alpha}}}^\textrm{me}$, respectively. These surface polarizability tensors relate the polarization surface-density vectors ${\bf P}$ and ${\bf M}$ to the incident electric and magnetic fields $\_E^\textrm{i}$ and $\_H^\textrm{i}$ through the constitutive relations~\cite{Serdyukov}
\e  \_P={\bar{\bar{\alpha}}}^\textrm{ee}\cdot \_E^\textrm{i} +{\bar{\bar{\alpha}}}^\textrm{em} \cdot \_H^\textrm{i}, \l{Pdef} \f
\e  \_M={\bar{\bar{\alpha}}}^\textrm{me} \cdot \_E^\textrm{i} +{\bar{\bar{\alpha}}}^\textrm{mm} \cdot \_H^\textrm{i}.\l{Mdef} \f
The superscript ``i'' denotes   ``incident'' fields, and the effective surface polarizability tensors take into account of the effect of multiple reflections inside the surface, i.e., of all the interactions among all the inclusions with themselves. The main goal in our synthesis problem is to find the required effective polarizability tensors ${\bar{\bar{\alpha}}}$ for a desired wave manipulation. We note that in some other studies, the impedance or susceptibility tensors are the main unknowns of the sysntehis problem~\citep{Achouri,Asadchy,Asadchy1,Epstein1,Epstein2}, versus the effective polarizability tensors as in this current investigation.\\
Similarly to what was done for planar metasurfaces~\cite{Achouri,Albooyeh3,Albooyeh31}, we start from the boundary conditions for cylindrical metasurfaces which read (see Eqs. (4.3a)--(4.3d) in Ref.~\cite{Idemen_book} and also Appendix~\ref{conditions} for a detailed derivation and the constitutive relations)
\e E_{z}^{+}-E_{z}^{-}=-\frac{1}{{{{\epsilon_0}}}}\frac{\partial {{{P}_{\rho }}}}{\partial z}+j\omega {{M}_{\phi }},\l{BC1} \f
\e E_{\phi}^{+}-E_{\phi }^{-}=-\frac{1}{{a}{{\epsilon_0}}}\frac{\partial {{{P}_{\rho }}}}{\partial \phi }-j\omega {{M}_{z}},\l{BC2} \f
\e H_{z}^{+}-H_{z}^{-}=-\frac{1}{{{\mu_0}}}\frac{\partial {{{M}_{\rho }}}}{\partial z}-j\omega {{P}_{\phi }},\l{BC3} \f
\e H_{\phi }^{+}-H_{\phi }^{-}=-\frac{1}{{a}{{\mu_0}}}\frac{\partial {{{M}_{\rho }}}}{\partial \phi }+j\omega {{P}_{z}},\l{BC4} \f
Here the $+$ and $-$ signs in the superscripts correspond to the total fields outside and inside of the metasurface boundary (i.e. $\rho=a$), respectively, and the implicit time dependence $\textrm{e}^{j\omega t}$ is assumed, with $\omega$ being the angular frequency. Moreover, $E_{\phi,z}$, $H_{\phi,z}$, $P_{\rho,\phi,z}$, and $M_{\rho,\phi,z}$ are correspondingly the field and surface polarization density components in cylindrical coordinates. Furthermore, ${{\mu_0 }}$  and ${{\epsilon_0 }}$ are the free space permeability and permittivity, respectively. Indeed, Eqs. ~\r{BC1}--\r{BC4} relate the jump of the tangential components of the electric and magnetic fields across the metasurface sheet to the required surface electric and magnetic polarization densities. The right hand sides of Eqs.~\r{BC1} and~\r{BC2} represent the {\it total} equivalent tangential magnetic surface polarization densities contributing to the discontinuity of the tangential electric field, whereas the right hand sides of Eqs.~\r{BC3} and~\r{BC4} represent the {\it total} equivalent tangential electric surface polarization densities contributing to the discontinuity of the tangential magnetic field~\cite{felsen,AlbooyehEqui,Albooyeh_Nromal}.\\
Next, for simplicity we only consider cases with vanishing normal polarization components $P_{\rho}$, and $M_{\rho}$ in the boundary conditions~\r{BC1}--\r{BC4} to simplify the analysis and synthesis without loosing any freedom in the manipulation of a desired electromagnetic field~\cite{disc1}. Indeed, based on the Huygens principle, or more precisely from the equivalence theorem~\cite{Harrington}, it can be shown that a desired field in a given volume can be fully engineered knowing the {\it tangential} (with respect to the metasurface surface) surface polarization components on the metasurface sheet~\cite{AlbooyehEqui,Albooyeh_Nromal}. As a result, the boundary conditions for the total field ~\r{BC1}--\r{BC4} simplify to
\e E_{z}^{+}-E_{z}^{-}=+j\omega {{M}_{\phi }},\l{BBC1} \f
\e E_{\phi}^{+}-E_{\phi }^{-}=-j\omega {{M}_{z}},\l{BBC2} \f
\e H_{z}^{+}-H_{z}^{-}=-j\omega {{P}_{\phi }},\l{BBC3} \f
\e H_{\phi }^{+}-H_{\phi }^{-}=+j\omega {{P}_{z}}.\l{BBC4} \f
In the next step, since cylindrical structures are inherently periodic with respect to the azimuthal coordinate $\phi$, we express the general form of the electromagnetic fields (inside and outside of the closed metasurface boundary) and the surface polarization densities on the metasurface as Fourier series
\e \mathbf{E}(\rho ,\phi ,z)=\sum\limits_{n=-\infty }^{n=\infty }{{{\mathbf{E}}_{n}}(\rho,z )}{{e}^{jn\phi }},\l{Ef} \f
\e \mathbf{H}(\rho ,\phi ,z)=\sum\limits_{n=-\infty }^{n=\infty }{{{\mathbf{H}}_{n}}(\rho,z )}{{e}^{jn\phi }},\l{Hf} \f
\e \mathbf{P}(\phi ,z)=\sum\limits_{n=-\infty }^{n=\infty }{{{\mathbf{P}}_{n}}(z)}{{e}^{jn\phi }},\l{Pf} \f
\e \mathbf{M}(\phi ,z)=\sum\limits_{n=-\infty }^{n=\infty }{{{\mathbf{M}}_{n}}(z)}{{e}^{jn\phi }},\l{Mf} \f
respectively, where ${{\mathbf{E}}_{n}}(\rho,z )$ and ${{\mathbf{H}}_{n}}(\rho,z )$ are the coefficients of the electromagnetic fields in the Fourier series whereas ${{\mathbf{P}}_{n}}(z)$ and ${{\mathbf{M}}_{n}}(z)$ are the coefficients corresponding to the electric and magnetic surface polarization densities on the metasurface sheet.\\
Before the next step and with the goal of practical implementation of the problem, we add one more level of simplification to our problem. That is, we consider no variation of the fields and polarization densities along the axial direction (here, $z$-direction) as we provide two examples in this study. It should be noted that the general synthesis of three-dimensional problems is a simple generalization of the $z$-invariant case studied here. Therefore, applying~\r{Ef}--\r{Mf} into the boundary conditions~\r{BBC1}--\r{BBC4}, assuming no $z$ variation, and considering $\frac{\partial }{\partial \phi }\to jn$, the boundary conditions for each Fourier coefficient due to the orthogonality of different $n$-indexed harmonics of the total field at $\rho=a$ read

\e E_{z,n}^{+}-E_{z,n}^{-}=+j\omega {{M}_{\phi,n }},\l{BCf1} \f
\e E_{\phi,n}^{+}-E_{\phi,n}^{-}=-j\omega {{M}_{z,n}},\l{BCf2} \f
\e H_{z,n}^{+}-H_{z,n}^{-}=-j\omega {{P}_{\phi,n }},\l{BCf3} \f
\e H_{\phi,n}^{+}-H_{\phi,n}^{-}=+j\omega {{P}_{z,n}}.\l{BCf4} \f
Equations~\r{BCf1}--\r{BCf4} are the final forms of the boundary conditions with the assumed simplifications and will be exploited hereafter.

\section{Incident fields}\label{incindentfield}
We consider two different scenarios: the illumination source either outside or inside the metasurface boundary [see Fig.~\ref{geom}].
Based on our examples in Sec.~\ref{examples}, we consider the plane wave excitation for the case of outside illumination whereas we discuss the infinite line source for the case of illumination from inside. However, we note that it is possible to consider any other excitation types based on the same analysis and procedure that follows.

\subsection{Plane wave illumination from outside}\label{illoutside}
For generality, we consider the superposition of two plane waves as external illumination: one with a $z$-polarized electric field i.e.,\e
\_E^\textrm{i}_{\rm TM}= \hat{\_{z}} E_0\textrm{e}^{-j\beta_{+} x},
\l{Eiplane}
\f
and the other with a $z$-polarized magnetic field i.e., 
\e
\_H^\textrm{i}_{\rm TE}= \hat{\_{z}} H_0\textrm{e}^{-j\beta_{+} x},
\l{Hiplane}
\f
which are traveling along the $+x$ direction (normal to the metasurface axis $z$) and impinging on the cylindrical metasurface. While the former is called  $\rm TM^z$--transverse magnetic-- polarization, the latter is called $\rm TE^z$--transverse electric [see Fig.~\ref{geom}]. Here, $\beta_{+}$ is the propagation constant in the medium outside of the metasurface boundary, whereas ${{E}_{0}}$ and $H_0$ are the incident electric and magnetic field amplitudes, respectively. Any plane wave which is propagating along the $x$ direction is decomposable into these two modes. For example, let us consider a plane wave hitting the metasurface as shown in Fig.~\ref{geom}, with its electric field component making an angle $\theta_0$ with the $z$-axis. In this case, we have $\eta_+ H_0=E_0 \tan\theta_0$, where $\eta_{+}=\sqrt{\mu_+/\epsilon_+}$ is the intrinsic wave impedance of the corresponding medium.\\
Next, to solve the problem of scattering from cylindrical structures, it is convenient to express the fields of Eqs.~\r{Eiplane} and~\r{Hiplane} in terms of  the cylindrical wave functions. It can be shown that the  $\rm TM^z$ incident plane wave with the electric field~\r{Eiplane} can be expressed as~\cite{balanis}
\e
\begin{array}{l}
\displaystyle
\_E^\textrm{i}_{\rm TM}= \hat{\_{z}}~{E_0}~\sum\limits_{n=-\infty }^{n=\infty } {j^{-n}}{{J}_{n}}({{\beta }_{+}}{{\rho }}) {{e}^{jn\phi }}, \end{array}\l{out_in_E}
\f
and that with $\rm TM^z$ polarization has magnetic field  \r{Hiplane} represented as
\e
\begin{array}{l}
\displaystyle
\_H^\textrm{i}_{\rm TE}= \hat{\_{z}}~{H_0}~\sum\limits_{n=-\infty }^{n=\infty } {j^{-n}}{{J}_{n}}({{\beta }_{+}}{{\rho }}) {{e}^{jn\phi }}. \end{array}\l{out_in_H}
\f
Here ${J}_{n}$ is the $n$-th order Bessel function and $\rho$ is the radial position. As a result, the total incident electric field which is the superposition of both the TE and TM incident plane waves reads\e
\begin{array}{l}
\displaystyle
\_E^\textrm{i}=\frac{1}{j\omega \epsilon_+}\nabla\times\_H_{\rm TE}^{\rm i}+\_E^\textrm{i}_{\rm TM}
\vspace*{.2cm}\\\displaystyle
\hspace*{0.4cm}
=\sum\limits_{n=-\infty }^{n=\infty } j^{-n}\left[{\eta_+H_0} \left({\hat{\boldsymbol{\rho}}} ~{n{J}_{n}({{\beta }_{+}}{{\rho}}) \over {{\beta }_{+}}\rho}
+ {\hat{\boldsymbol{\phi}}}~{j}{{J}^{\prime}_{n}}({{\beta }_{+}}{{\rho }})\right) 
\right.\vspace*{.2cm}\\\displaystyle
\hspace*{0.6cm}\left.
+\hat{\_{z}}~{E_0}{{J}_{n}}({{\beta }_{+}}{{\rho }}) \right] {{e}^{jn\phi }}, \end{array}\l{out_inc}
\f
where ${\hat{\boldsymbol{\rho}}}$, $\hat{\boldsymbol{\phi}}$, and $\hat{\_z}$ are the unit vectors in cylindrical coordinates and ${{J}^{\prime}_{n}}$ is the derivative of the $n$-th order Bessel function with respect to the argument ${{\beta }_{+}}{{\rho }}$.
\subsection{Line source illumination from inside}\label{illinside}
Here we consider the superposition of two infinitely extended electric and magnetic line sources located at the center of the cylindrical metasurface as excitation source. The current amplitude of the electric and magnetic line sources are assumed to be $I_e$ and $I_m$, respectively. Similarly to the previous case of plane wave illumination from outside, here the electric line source creates a $\rm TM^z$ field while the magnetic one generates a  $\rm TE^z$ field. The electric and magnetic fields of the two kinds of sources are, respectively, given by~\cite{balanis}
\e
\_E^\textrm{i}_{\rm TM}= -\hat{\_{z}} {{I}_{e}} \frac{\beta _{-}^{2}}{4\omega {{\epsilon }_{-}}}H_{0}^{(2)}({{\beta }_{-}}{{\rho }}),
\l{E_line}
\f
and\e
\_H^\textrm{i}_{\rm TE}= -\hat{\_{z}} {{I}_{m}} \frac{\beta _{-}^{2}}{4\omega {{\mu }_{-}}}H_{0}^{(2)}({{\beta }_{-}}{{\rho }}),
\l{H_line}
\f
where $H_{0}^{(2)}$ is the $0$-th order Hankel function of the second kind. Therefore the total incident electric field is
\e
\begin{array}{l}
\displaystyle
\_E^\textrm{i}=\frac{1}{j\omega \epsilon_-}\nabla\times\_H_{\rm TE}^{\rm i}+\_E^\textrm{i}_{\rm TM}
\vspace*{.2cm}\\\displaystyle
\hspace*{0.4cm}
=-{\beta_-\over 4}\left[ {\hat{\boldsymbol{\phi}}}~j{{I}_{m}} {H_{0}^{(2)}}^{\prime}({{\beta }_{-}}{{\rho }})+\hat{\_{z}}~\eta_{-} {{I}_{e}} {H_{0}^{(2)}}({{\beta }_{-}}{{\rho }}) \right]. \end{array}\l{in_inc}
\f
Note that the sources could be located also elsewhere inside the cylindrical metasurface boundary e.g. at ${\boldsymbol{\rho}}^{\prime}$ away from the origin, and in this case one would need to apply the substitution $\rho=|{\boldsymbol{\rho}}-{\boldsymbol{\rho}}^{\prime}|$, where ${\boldsymbol{\rho}}$ is the observation point.\\
In the next steps, we present the general forms of the total fields due to the induced sources on the metasurface boundary, i.e., due to the secondary sources.

\section{Fields in cylindrical coordinates}\label{scattfields}
Generally, the fields inside ($\_E_{\rm in}$) and outside ($\_E_{\rm out}$) of the cylindrical metasurface boundary generated by the induced currents (scattered fields) can be expressed as
\e
\begin{array}{l}
\displaystyle
\_E_{\rm in}=  \sum\limits_{n=-\infty }^{n=\infty } \left[{\hat{\boldsymbol{\rho}}}~ n{{b_n}}{{J}_{n}({{\beta }_{-}}{{\rho }})\over {{\beta }_{-}}\rho}
\right.\vspace*{.2cm}\\\displaystyle
\hspace*{0.0cm}\left.
+ {\hat{\boldsymbol{\phi}}}~ j{{b_n}}{{J}^{\prime}_{n}}({{\beta }_{-}}{{\rho }}) +\hat{\_{z}}~{{a_n}}{{J}_{n}}({{\beta }_{-}}{{\rho }}) \right] {{e}^{jn\phi }}, \end{array}\l{out_scat_in}
\f
and
\e
\begin{array}{l}
\displaystyle
\_E_{\rm out}=  \sum\limits_{n=-\infty }^{n=\infty } \left[{\hat{\boldsymbol{\rho}}}~n{{d_n}}{{H}^{(2)}_{n}({{\beta }_{+}}{{\rho }})\over {{\beta }_{+}}\rho}
\right.\vspace*{.2cm}\\\displaystyle
\hspace*{0.0cm}\left.
+ {\hat{\boldsymbol{\phi}}}~ j{{d_n}}{{H}^{(2)}_{n}}'({{\beta }_{+}}{{\rho }}) +\hat{\_{z}}~{{c_n}}{{H}^{(2)}_{n}}({{\beta }_{+}}{{\rho }}) \right] {{e}^{jn\phi }}, \end{array}\l{out_scat_out}
\f
respectively. Note that the scattered field expressions in Eqs.~\r{out_scat_in} and~\r{out_scat_out} do not include the fields produced by either the inner line source or the incident plane wave, and only consider the fields generated by the induced currents on the cylindrical metasurface, i.e., secondary sources. Here ${{H}^{(2)}_{n}}$ and ${{H}^{(2)}_{n}}'$ are the $n$-th order Hankel function of the second kind and its derivative with respect to the argument, respectively. Moreover, the first two components i.e., $\rho$ and $\phi$ components in Eqs.~\r{out_scat_out} and~\r{out_scat_in} correspond to the $\rm TE^z$ polarized fields while the $z$ component  corresponds to the $\rm TM^z$ polarized fields. Furthermore, due to the nonsingular nature of the scattered fields inside and outside we have expanded the fields in terms of the suitable Bessel and Hankel functions, respectively. The magnetic fields ($\_H_{\rm out}$ and $\_H_{\rm in}$) can be calculated using Maxwell's equations. Note that ${{a}_{n}}$ and ${{b}_{n}}$ are coefficients of Bessel-Fourier series representing standing waves inside the metasurface boundary for {$\rm TM^z$ and $\rm TE^z$} waves, respectively, whereas ${{c}_{n}}$ and ${{d}_{n}}$  are the coefficients which represent the {$\rm TM^z$ and $\rm TE^z$} waves outside the metasurface, correspondingly.\\
At this stage we have all tools to synthesize a specific and desired electromagnetic wave. To summarize, in order to synthesize a desired wave profile:
\begin{enumerate}
\item We first expand it in terms of the cylindrical harmonics as given by Eqs.~\r{out_scat_in} and~\r{out_scat_out} to obtain the unknown coefficients ${{a}_{n}}$, ${{b}_{n}}$, ${{c}_{n}}$, and ${{d}_{n}}$.

\item Next, by applying the given incident fields~\r{out_inc} or~\r{in_inc} and by exploiting the boundary conditions~\r{BCf1}--\r{BCf4}, we find the required coefficients $\_P_n$ and $\_M_n$ that provide the required polarization densities $\_P$ and $\_M$.

\item Finally, when the polarization densities are found, the required effective polarizability tensors ${\bar{\bar{\alpha}}}^\textrm{ee}$, ${\bar{\bar{\alpha}}}^\textrm{mm}$,${\bar{\bar{\alpha}}}^\textrm{em}$, and ${\bar{\bar{\alpha}}}^\textrm{me}$ are retrieved from the constitutive relations~\r{Pdef} and~\r{Mdef}.

\end{enumerate}
As a final remark, note that each polarization density vector has two tangential components $P_\varphi, P_z$ and $M_\varphi, M_z$  providing a total of four functional equations. We recall that for simplicity we consider metasurfaces that express only tangential polarization densities and exclude those metasurfaces with normal polarization densities $P_\rho$ and $M_{\rho}$. However, each polarizability tensor in Eqs. \r{Pdef} and \r{Mdef} has four tangential polarizability components $\alpha_{\varphi\varphi}$, $\alpha_{\varphi z}$, $\alpha_{z\varphi}$, $\alpha_{zz}$. Therefore, the solution to the synthesis of a metasurface is not unique since there are two polarization vectors and four polarizability tensors, i.e., there are four equations with $16$ complex unknown polarizability components.

\section{Illustrative examples}\label{examples}
We present three representative examples to demonstrate the application of the proposed method and to clarify the synthesis approach. Other manipulations of electromagnetic waves in cylindrical coordinates shall be possible using a similar approach.\\
In the first example we consider the application of a metasurface as a scattering cancellation device~\cite{Chen,Schurig,Padooru} for a cylindrical perfect electric conductor (PEC) when the system is excited by a plane wave. In the second example we investigate an application similar to that of the first example with the only difference that the PEC cylinder is replaced by a dielectric cylinder. The main advantage in using a metasurface for the realization of a scattering cancellation device compared to previous methods such as transformation optics (TO)~\cite{Schurig,Yi} and transmission line method~\cite{Tretyakov1} is that one does not need to use bulky materials with exotic properties which inevitably take a large space and/or contain high losses. Instead, a simple electromagnetically thin composite surface which is easier to fabricate and take less space is exploited.\\
In the last example we consider a metasurface that surrounds a line source that creates a fictitious line source outside the cylindrical metasurface boundary. Following Ref.~\onlinecite{Yi}, we call this fictitious source the illusory source since an observer outside the metasurface boundary sees only a line source which is displaced with respect to the original source. The realization of an illusion device using the available approaches such as TO requires the source to be located in a bulky complex medium which makes it mainly impractical. However, in our approach, the real source is simply located in free space and we surround it with a thin metasurface. Therefore, the advantages of our approach compared to the previous approaches for the design of electromagnetic devices for cylindrical structures is two-fold; the proposed approach is both practical and general.\\

\subsection{Scattering cancellation from a PEC cylinder}\label{Excloak}
\begin{figure}[h!]
\centering
 \epsfig{file=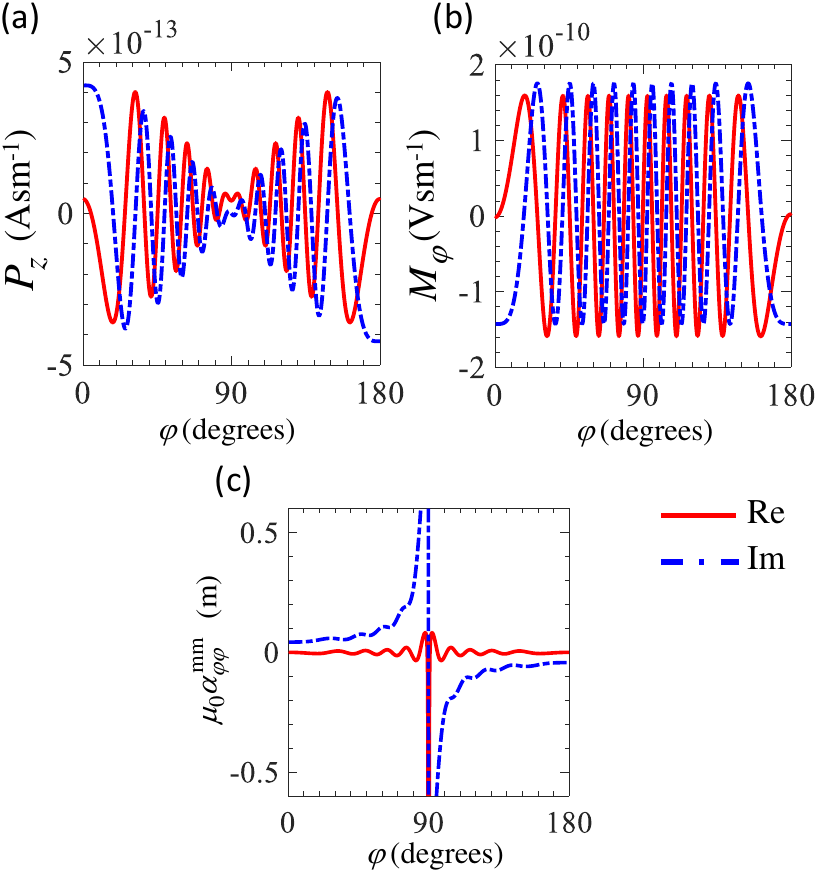, width=0.99\linewidth}  
  \caption{ Induced surface (a) electric $P_z$ and (b)  magnetic $M_\varphi$ polarization densities of Eqs.~\r{P_PEC} and~\r{M_PEC} required for scattering cancellation. (c) The corresponding effective magnetic  surface polarizability $\alpha_{\varphi\varphi}^{\rm mm}$ desired for scattering cancellation from a PEC cylinder with radius $a=5\lambda$. The values in plots are similarly repeated for $180^\circ \leq \phi \leq 360^\circ$}\label{polcl1}
\end{figure}

\begin{figure}[h!]
\centering
 \epsfig{file=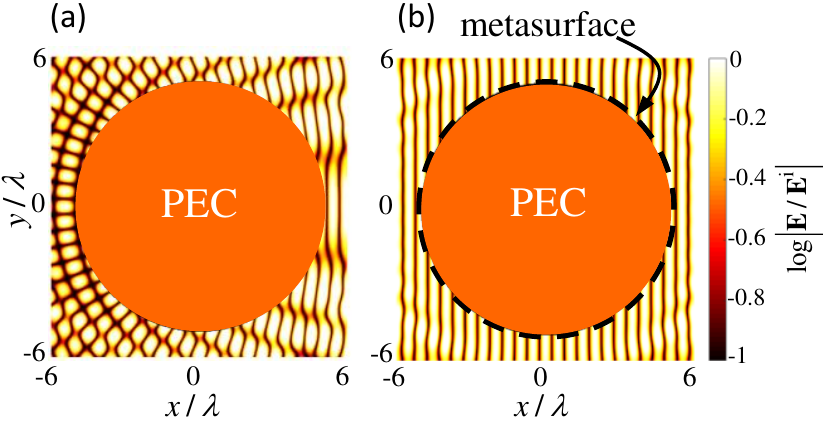, width=0.99\linewidth}  
  \caption{The normalized (to the incident field amplitude $\_E^{\rm i}$) electric field distribution around the proposed PEC cylinder illuminated by a plane wave in (a) the absence and (b) the presence of the proposed scattering cancellation cylindrical metasurface. The metasurface is shown with a black dashed line.}\label{cl1}
\end{figure}

In this first example we consider a PEC cylinder with radius $a=5\lambda$ ($\lambda$ is the wavelength of the excitation field) located in free space which is extended infinitely along its axis (here the $z$-axis) and is illuminated by a $z$-polarized plane wave propagating along the $x$-axis as given by Eq.~\r{Eiplane} (i.e., $\rm TM^z$ incidence) with $E_0=1~{\rm V}{\rm m}^{-1}$ and $ {\omega/{(2\pi)}} = 1~{\rm GHz} $. Our goal is to find a proper metasurface which covers the PEC cylinder and suppresses the scattered electromagnetic field by the PEC. In order to cancel the scattered field, the total electric and magnetic fields must satisfy $\_E^{+}=\_E^{\rm i}$, $\_H^{+}=\_H^{\rm i}$ outside the metasurface, and inside the PEC, $\_E^{-}=0$ and $\_H^{-}=0$, respectively. Therefore, by applying the boundary conditions~\r{BCf1}--\r{BCf4} one finds the required surface polarization density coefficients $\_P_n$ and $\_M_n$ at the boundary. Hence, the desired surface electric and magnetic polarization densities $\_P$ and $\_M$ at the boundary are obtained from Eqs.~\r{Pf} and~\r{Mf} as \e \_P=-\hat{\_z} {E_0\over \omega\eta_0}\sum\limits_{n=-\infty }^{n=+\infty} { J^{\prime}_{n}(\beta_0 a)  j^{-n} e^{jn\phi} } ,\l{P_PEC} \f 
and
\e \_M=\hat{\boldsymbol{\phi}}{E_0\over j\omega}\sum\limits_{n=-\infty }^{n=+\infty} { J_{n}(\beta_0 a)  j^{-n} e^{jn\phi} }. \l{M_PEC} \f  
Here, $\eta_0=\sqrt{\mu_0/\epsilon_0}$ is the intrinsic impedance of free space and $\beta_0=\omega\sqrt{\mu_0\epsilon_0}$ is the propagation constant of the incident wave (i.e., $\beta_0 a=10\pi$ in this example). Note that for the proposed $\rm TM^z$ incidence we have $\theta_0=0$ [see Fig.~\ref{geom}]. Indeed for this special case, the required $\phi$-component of $\_P$ and the $z$-component of $\_M$ vanish, i.e., $P_z$ and $M_\phi$ are the sufficient surface polarization density components at the boundary to achieve the scattering cancellation. Similarly, for $\rm TE^z$ incidence as in Eq.~\r{Hiplane}, the surface polarization density components $P_\phi$ and $M_z$ are sufficient to suppress the scattered fields. Note that based on the reciprocity theorem the scattered fields of any impressed surface electric polarization density on the surface of a PEC is zero, therefore, the obtained surface electric polarization density $\_P$ in Eq.~\r{P_PEC} represents the induced current on the PEC surface due to the incident plane wave. However, the magnetic polarization density $\_M$  at the boundary is an impressed polarization density provided by the metasurface to cancel the scattered fields by the PEC. The surface polarization densities of Eqs.~\r{P_PEC} and~\r{M_PEC} that synthesize the scattering cancellation of a PEC cylinder are depicted in Fig.~\ref{polcl1}(a) and (b). 
\\
The last step is to retrieve the required effective polarizability tensors from Eqs.~\r{Pdef} and~\r{Mdef} that realize the polarization density $\_M$ obtained in Eq.~\r{M_PEC}. However, in Eq.~\r{Mdef} the number of equations are less than the number of unknown polarizability components. Therefore, the solution to this synthesis problem is not unique since there are multiple polarizability tensors' sets which lead to the same magnetic polarization density $\_M$, hence, to the desired electromagnetic field. Here, as previously mentioned we simplify the synthesis procedure by neglecting the polarizability components normal to the metasurface plane (i.e., $\rho$-components). Moreover, for the sake of simplicity, we avoid considering realizations that possess bianisotropic polarizabilities ${\bar{\bar{\alpha}}}^\textrm{em}$ and ${\bar{\bar{\alpha}}}^\textrm{me}$. Furthermore, we consider realizations with vanishing cross-component polarizability components, i.e., with zero off-diagonal terms in the polarizability tensor ${\bar{\bar{\alpha}}}^\textrm{mm}$. To summarize, (i) $\_P$ is the induced surface polarization density on the PEC surface and not a surface polarization density impressed on the metasurface; and (ii) the surface polarization density $M_\varphi$ given by Eq.~\r{M_PEC} to provide scattering cancellation is obtained with a realization of the metasurface with effective surface polarizability component $ \alpha_{\varphi\varphi}^{\rm mm}$ shown in Fig.~\ref{polcl1}(c). As it is clear, the effective polarizability component $ \alpha_{\varphi\varphi}^{\rm mm}$ varies with the angle, i.e., we require a spatially varying magnetic current to cancel the scattering from a PEC cylinder by using a metasurface. For a $\rm TE^z$ plane wave incidence, the solution of canceling the scattered field is dual to the one just mentioned and the metasurface would have only surface polarizability component $\alpha_{zz}^{\rm mm}$. Figures~\ref{cl1}(a) and (b) demonstrate the electric field distributions in the absence and presence of the synthesized metasurface, respectively, by using full-wave simulations based on the finite element method~\cite{COMSOL}.
It is clear from this Figure that the by placing the synthesized metasurface around the PEC cylinder, the cylinder's scattered fields have perfectly canceled, in other words the metasurface acts as a cloaking device for a PEC cylinder. 
We next examine the cancellation of the scattering (i.e., cloaking) of a dielectric cylinder by covering it by a metasurface to further demonstrate the capability of our proposed general approach.

\subsection{Scattering cancellation from a dielectric cylinder}\label{Excloak2}
\begin{figure}[h!]
\centering
 \epsfig{file=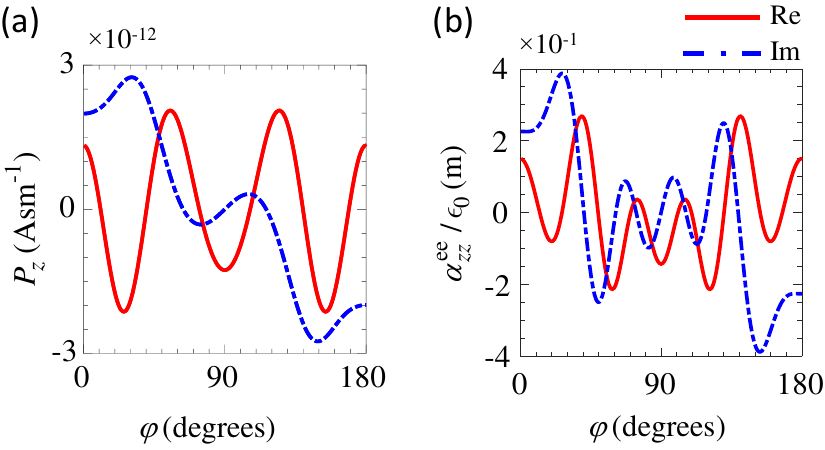, width=0.99\linewidth}  
  \caption{(a) The required polarization density for the example of scattering cancellation metasurface for the proposed dielectric cylinder. (b) The corresponding effective electric polarizability of the designed metasurface. The dielectric radius and relative permittivity are $a=\lambda$ and $\epsilon_r=10$, respectively.}\label{polcl2}
\end{figure}

\begin{figure}[h!]
\centering
 \epsfig{file=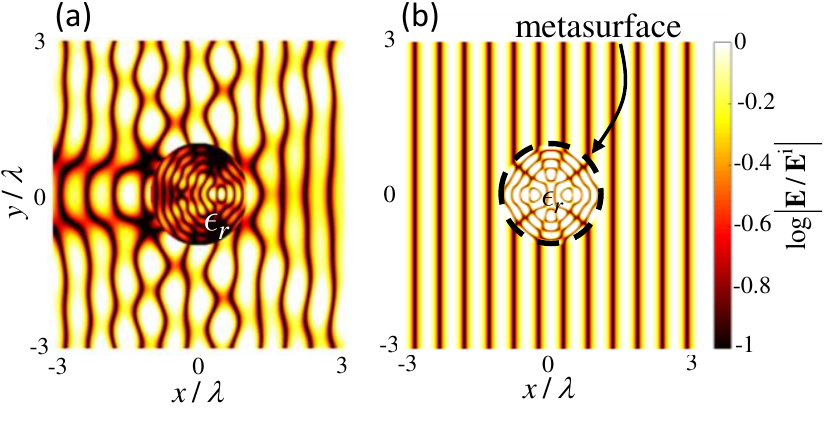, width=0.99\linewidth}  
  \caption{The normalized (to the incident field amplitude $\_E^{\rm i}$) electric field distribusions in (a) the absence and (b) the presence of the synthesized scattering calncellation cylindrical metasurface in the case of a dielectric cylinder. The metasurface is shown with a black dashed line.}\label{cl2}
\end{figure}

We consider now the problem of scattering cancellation from an infinitely long dielectric cylinder with relative permittivity $\epsilon_r=10$ and radius $a= \lambda$ located in free space and illuminated by an electromagnetic plane wave as in Eq.~\r{Eiplane} as in the previous case. There is a major difference between this case and the previous one with a PEC cylinder. In the previous example, the electromagnetic fields $\_E^{-}$ and $\_H^{-}$ inside the cylinder were required to be zero. Accordingly, based on the boundary conditions~\r{BCf1}--\r{BCf4}, providing scattering cancellation enforces both electric and magnetic polarization densities $\_P$ and $\_M$ to be present at the boundary of the PEC ~\cite{Kwon,albooyeh_abs, radi_abs1,radi_abs2,alaee_abs}. In the present case with a dielectric cylinder, there are infinite solutions that solve the scattering cancellation problem depending on the precise form which the equivalent principle is applied. Among them, we consider the one which can be realized when we impose that only the electric polarization density $\_P$ is non vanishing. Therefore, assuming the magnetic polarization density $\_M$ to be zero in Eqs.~\r{BCf1} and~\r{BCf2}, the electromagnetic field inside the dielectric cylinder can be easily obtained. Based on the considered incident polarization, the electric field inside the metasurface has a $z$-component only. Moreover, to have no scattered fields out of the metasurface (i.e., $\_E_{\rm out}=0$), we apply the boundary condition~\r{BCf1} using Eq.~\r{out_in_E} to obtain the only surviving coefficient of Eq.~\r{out_scat_in} as $a_n=E_0 j^{-n} {J_n(\beta_0 a)}/\left({J_n(\beta_0\sqrt{\epsilon_r} a)}\right)$. As a result, the required surface polarization density $\_P$ is obtained from the boundary condition~\r{BCf4} and using the Maxwell-Faraday equation $\nabla \times \_E^{\pm}=-j\omega\mu_0\_H^{\pm}$ at the metasurface boundary (i.e., $\rho=a$), leads to

\begin{equation}
\begin{array}{l}
\displaystyle
\_P=-\hat{\_z}{E_0\over \omega\eta_0}\sum\limits_{n=-\infty }^{n=+\infty} j^{-n} e^{jn\phi} \left[J'_{n}(\beta_0 a)
\right.\vspace*{.2cm}\\
\displaystyle
\hspace*{1.5cm}\left.
-{\sqrt{\epsilon_r}J_{n}(\beta_0 a)\over J_{n}(\beta_0\sqrt{\epsilon_r}a)}J'_{n}(\beta_0\sqrt{\epsilon_r} a)  \right].
\end{array}
\l{P_DIEL}
\end{equation}
The longitudinal polarization component is plotted in Fig.~\ref{polcl2}(a) as a function of the azimuthal angle $\varphi$. 
Next we derive the polarizabilities that provide the required polarization density. For this simple case only one kind of polarizability is sufficient. Indeed, we assume that the bianisotropic polarizabilities ${\bar{\bar{\alpha}}}^\textrm{em}$ and ${\bar{\bar{\alpha}}}^\textrm{me}$ are absent as well as the cross-component polarizabilities. Then, by using the constitutive relation~\r{Pdef} the required effective electric polarizability $\alpha_{zz}^{\rm ee}$ is derived and depicted in Fig.~\ref{polcl2}(b). The electric field distributions (assuming plane wave incidence) in the absence and presence of the designed metasurface are, respectively,  demonstrated in Fig.~\ref{cl2}(a) and (b) by using full-wave simulations.
As it is clear from the figure, when the metasurface is present the scattered field is suppressed outside the metasurface boundary. That is, outside the metasurface we observe only the propagating incident plane wave.

In both the last examples, as seen in Figs.~\ref{polcl1} and~\ref{polcl2}, the imaginary part of the required effective polarizability tensors is negative at some azimuthal positions and positive for others. Therefore, one may deduce that the desired metasurface consist of active elements. However, as discussed in~\cite{Asadchy,Asadchy1}, the pointing vector of the wave can switch the direction locally at the metasurface plane and when its direction is opposite to the incident wave, the polarizabilities can be negative. This implies the possibility of realization of the synthesized structures using only passive particles.
 
\subsection{An illusory infinite line source}\label{Exillusion}
In this example the goal is to create an electromagnetic illusion that a source is moved somewhere else by surrounding the actual source with a metasurface. 
Studies on the design of optical devices based on TO to create optical illusions have been already performed in past recent years (see e.g. Ref.~\onlinecite{Yi}). However, besides proposing bulky complex media, there was also the restriction that both the radiation source and its illusion were located inside the engineered medium. These requirements make the realization of such devices difficult if not impractical and also less useful. However, as we propose here, the introduction of metasurfaces make the realization of such devices more convenient and practical. Here, the goal is to show how a cylindrical metasurface around a line source is able to modify the source fields to make it look like it is generated by a fictitious line source outside the metasurface boundary, translated from the original source that is not visible anymore. In summary the metasurface cloaks the original line source and generates an illusory line source at a different location.

Let us consider an infinitely long electric line source $I_e$ in free space, along the $z$-axis of a cylindrical coordinate system. The electric field of such current source is described by Eq.~\r{E_line}. The goal is to synthesize a cylindrical metasurface around this line source that creates a total field that looks like the one generated by a fictitious line source translated at {$({\rho}'>a,{\phi}')$}, for an observer outside the metasurface [see Fig. \ref{ill1}]. Moreover, the distance between the metasurface and the fictitious line source must be subwavelength due to energy conservation considerations. Indeed, the total flux of the Poynting vector through any closed surface around the illusory source must be zero if such surface does not enclose any part of the metasurface. This would be impossible if the fictitious line source is located at large distance from the metasurface. 

We assume the fictitious line source has the same amplitude $I_e$ as the original line source but is located at $\rho ={\rho }',\phi ={\phi}'$. By using the {\it addition theorem} (see e.g., Refs.~\cite{balanis,Harrington}) the electric field created by such a source, which corresponds to the desired total field outside the cylindrical metasurface, reads

\e {\_E^+}=-\hat{\_z}\frac{{{I}_{e}}\beta_0^{2}}{4\omega {{\epsilon_0}}}\sum\limits_{-\infty }^{+\infty }{{{J}_{n}}({{\beta_0 }}\rho^{\prime} )H_{n}^{(2)}({{\beta_0 }}{\rho }){{e}^{jn(\phi -{\phi}')}}}.\l{Eout_i1}\f
Comparing the above equation with Eq.~\r{out_scat_out} that represents the total scattered electric field outside the metasurface, one realizes that the only surviving coefficient in Eq.~\r{out_scat_out} is
\e
c_n=-\frac{I_e\beta_0^2}{4\omega\epsilon_0} J_n\left(\beta_0\rho'\right).\l{cn_illusion}
\f

The total electric field inside the cylindrical metasurface reads
\e {\_E^-}=-\hat{\_{z}}  \frac{I_e \beta_0^{2}}{4\omega {{\epsilon_0}}}H_{0}^{(2)}({{\beta_0}}{{\rho }})+\hat{\_z}\sum\limits_{-\infty }^{+\infty }{ a_n{{J}_{n}}({{\beta_0}}\rho ){{e}^{jn(\phi -{\phi}')}}},\l{Ein_i1}\f

which is the contribution of the field created by the original line source [see Eq.~\r{E_line}] plus the field created from the cylindrical metasurface [see Eq.~\r{out_scat_in}]. Note that the metasurface shall not create any cross component scattered field, i.e., the $\rho$- and $\phi$-components of the electric field simply vanish in Eq.~\r{out_scat_in}. Now we assume the fictitious source is located at ${\rho}'$ and ${\phi}'=0$. Moreover, we assume the observer is located at distances that satisfy $\rho > {\rho }'$. 
There are many possible realizations of such metasurface, and here we consider only realizations with non vanishing electric surface polarizations density and $\_M$ assumed to be zero. {As a result of the above considerations, and by using the electric fields~\r{Eout_i1} and~\r{Ein_i1} in the boundary condition~\r{BCf1}, the only surviving coefficients in Eq.~\r{out_scat_in} read
\e {a_n}=\left\{  \begin{matrix}
\displaystyle   -\frac{I_e\beta_0^2}{4\omega\epsilon_0} {H_0^{(2)}(\beta_0 a) \over J_0(\beta_0 a)} \left[ J_0\left(\beta_0\rho'\right)-1\right] & n=0 \\
   \\
\displaystyle   -\frac{I_e\beta_0^2}{4\omega\epsilon_0}\frac{H_n^{(2)}(\beta_0 a)}{J_n(\beta_0 a)} J_n\left(\beta_0\rho'\right)& n\neq 0
\end{matrix}  \right. .\l{an_illusion} \f
Finally, by applying the boundary condition~\r{BCf4} and by using the Maxwell-Faraday equation $\nabla \times \_E^{\pm}=-j\omega\mu_0\_H^{\pm}$ at the metasurface boundary (i.e., $\rho=a$), we obtain the required electric  polarization density $\_P$ for the desired fields given by~\r{Eout_i1} and~\r{Ein_i1} as
\begin{equation}
\begin{array}{l}
\displaystyle
\_P=-\hat{\_z} {I_e\beta_0\over 4\omega} \left({H_0^{(2)}}^\prime \left(\beta_0 a\right)-\sum\limits_{n=-\infty }^{n=+\infty}  e^{jn\phi} \left[J_{n}(\beta_0 \rho'){H_n^{(2)}}^\prime \left(\beta_0 a\right)
\right.\right.\vspace*{.2cm}\\
\displaystyle
\hspace*{5.6cm}\left.\left.
+{4a_n\over \omega\mu_0 I_e}  J'_{n}(\beta_0 a)  \right]\right),
\end{array}
\l{P_ILLU}
\end{equation}

where $a_n$ is given by Eq.~\r{an_illusion}.
This surface polarization density is used in the constitutive relation~\r{Pdef} to retrieve the required effective surface polarizability tensors. Figures~\ref{polill1}(a) and (b), respectively, illustrate the required surface polarization density and the needed effective electric surface polarizability for the proposed problem assuming $a=\lambda$ and $\rho'= 4\lambda/3 $.
\begin{figure}[h!]
\centering
 \epsfig{file=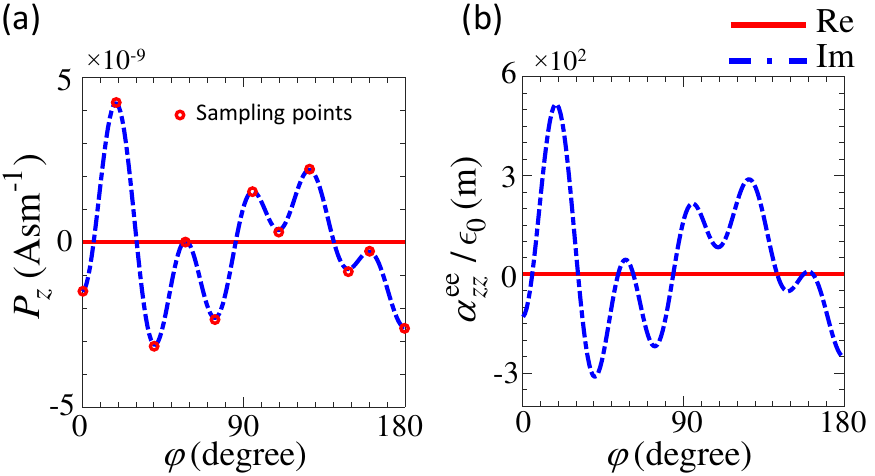, width=0.99\linewidth}  
  \caption{(a) The required polarization density for the metasurface example that creates an infinite illusory line source. (b) The corresponding effective electric polarizability of the designed metasurface.}\label{polill1}
\end{figure}
Similarly to what was done in the previous example, we again have considered a metasurface realization that does not involve cross-component and bianisotropic polarizabilities. Figures~\ref{ill1}(a) and (b), respectively, show the full-wave simulation results for the electric field distributions in the absence and the presence of the engineered metasurface.
\begin{figure}[h!]
\centering
 \epsfig{file=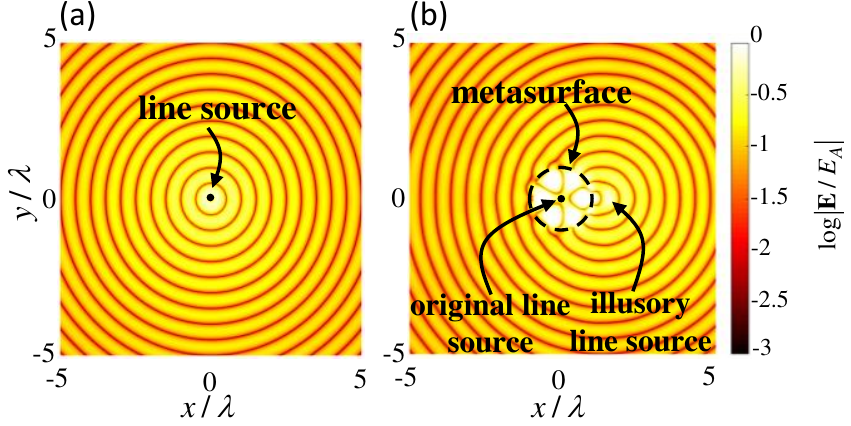, width=0.99\linewidth}  
  \caption{(a) The normalized [to $E_A = {I_e\beta_0^2/ \left(4\omega\epsilon_0\right)}$] electric field distribution around an infinitely long electric line source in vacuum. (b) The normalized electric field distribution of the original line source when located at the center of a metasurface that is synthesized to translate the perception of the position of the original line source for an observer outside the cylindrical metasurface. The matasurface radius is $a=\lambda$, the illusory source is translated to the position 
$\rho'= 4\lambda/3$ and $\phi'=0$, and the field appears arising from such source.}\label{ill1}
\end{figure}
It is clear from the field distributions that the original line source is cloaked and an illusory line source is observed at $4\lambda/3$ away from the original line source for an observer outside the cylinder with radius  $\rho'$, i.e., for $\rho > \rho'$. Such a device may find practical applications in stealth technology, wireless power transfer systems~\cite{Song}, and also sensing technologies.

\section{practical realization}\label{PR}
In all the above examples we have performed conceptual syntheses by considering ideal continuous current distributions, i.e., continuous surface polarization densities. However, a metasurface is practically composed of discrete elements that mimic such continuous currents. In this section, we present a practical approach to realize such metasurfaces with discrete elements, and apply it to the last example, i.e., the illusory line source. We start by taking samples along the azimuthal angular direction that corresponds to the extrema values of the desired continuous current distribution given in Fig.~\ref{polill1}(a). According to Fig.~\ref{polill1}(a) and considering the whole $360^\circ$ azimuth angle, we obtain $20$ sampling points. Note that only half of the whole $360^\circ$ angular span is shown in Fig.~\ref{polill1}(a) since the other half of the cylinder is symmetric to the one shown and hence it exhibits the same values. Next, each of these required currents can be realized by wires with  periodic set of loads located at the azimuthal sampling points. For simulation purposes we restrict our analysis to a single periodic unit as  shown in Fig.~\ref{ilusiondiscrete} and representing all the others by reflections of   two parallel perfectly conducting plates.
\begin{figure}[h!]
\centering
 \epsfig{file=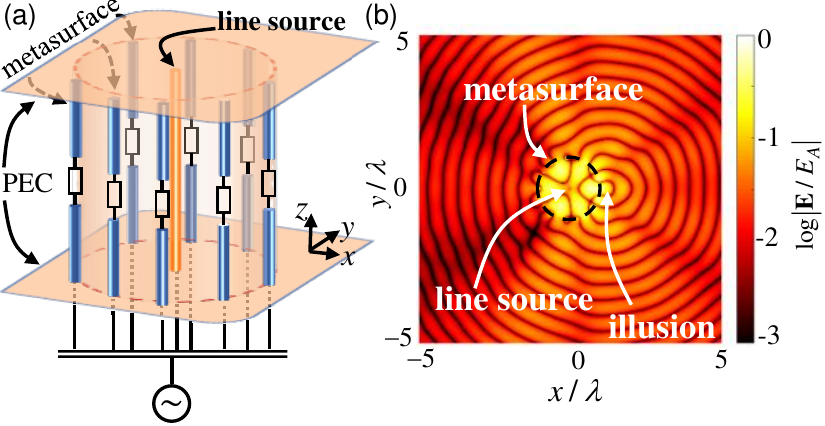, width=0.99\linewidth}  
  \caption{(a) The schematic of an exemplary practical realization of the discretized  metasurface that creates an illusory line source next to the original one that not visible. (b) The normalized [to $E_A = {I_e\beta_0^2/ \left(4\omega\epsilon_0\right)}$] electric field distribution for the structure given in (a).}\label{ilusiondiscrete}
\end{figure} 
It is noteworthy that the coupling between the dipoles in the parallel plate region are tuned by the loads to obtain the desired currents at the sampling points. In order to find the necessary load impedance values we first create a cylindrical array of $20$ identical short dipole antennas located at each sampling point. These points are given by the angular positions of the extrema values given in Fig.~\ref{polill1}(a). 
Each wire in the plates scatters and couples to all the others, modifying their currents and voltages. Therefore by looking at the voltage and current at each load we relate all these quantities with a  $20\times 20$ impedance matrix $\bar{\bar{Z}}$ in which the diagonal elements denote the self impedance of each antenna and the off-diagonal elements represent the mutual couplings between the antennas. The goal is to design the loads at each wire to ensure that the current distribution is similar to the one shown in Fig.~\ref{polill1}(a). For convenience we define the load impedance diagonal matrix $\bar{\bar{Z}}_{\rm L}$, which contains the load impedances $Z_{{\rm L}i}$ at each sampling points $\phi_i$, $i=1,..., 20$ shown in Fig.~\ref{imps}. The scattering problem is now posed as an algebraic problem as $\_V= \left[\bar{\bar{Z}}+\bar{\bar{Z}}_{\rm L}\right] \_I$, where $\_I$ is a $20$ element vector with its $i$-th element $I_i=j\omega a\_P(\phi-\phi_i)\cdot\hat{\_z}$ given by the effective surface electric  polarization density at the sampling point $\phi_i$ [Fig.~\ref{polill1}(a)] and $\_V$ is a voltage (per meter) vector applied to dipoles. Considering identical supplying voltage $1$ $({\rm V/m})$ for all the antennas, the required load impedances are given in Fig.~\ref{imps}.
\begin{figure}[h!]
\centering
 \epsfig{file=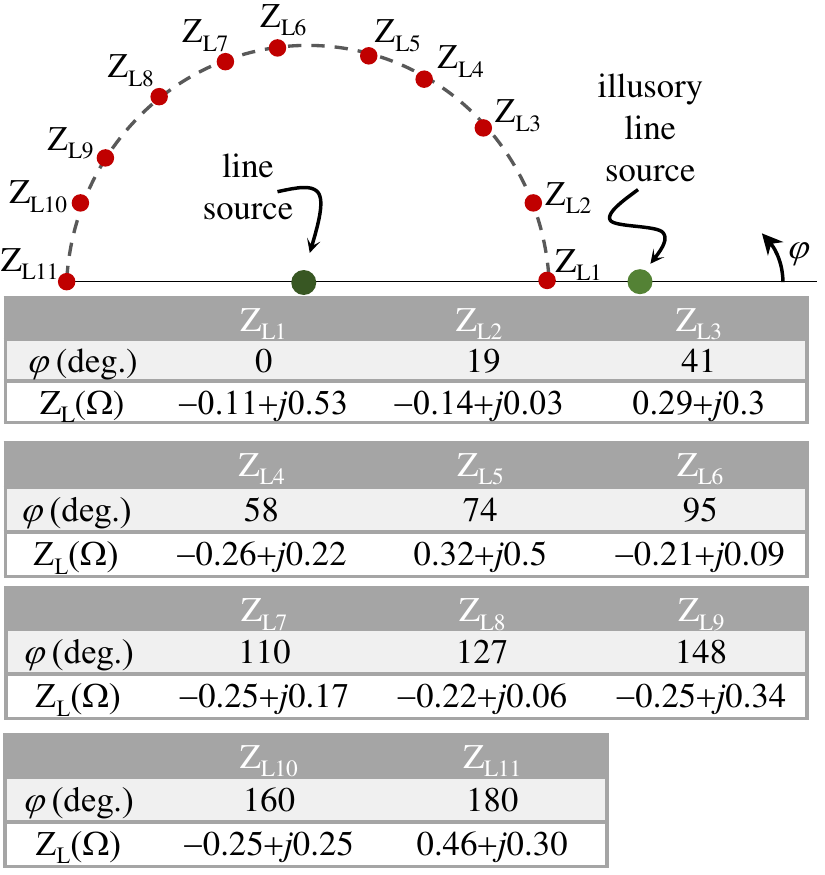, width=0.99\linewidth}  
  \caption{The relative positions and load impedance values of the dipoles in the proposed cylindrical metasurface that creates an illusory line source. Here, we show $11$ elements while we declare that elements number $2$ to $10$ are repeated symmetrically in the lower half circle.}
  \label{imps}
\end{figure}
Here we have provided the impedance values for half of the elements since the values for the other half is symmetrically distributed [see Fig.~\ref{imps}]. The electric field distribution of such discretized problem is depicted in Fig.~\ref{ilusiondiscrete}(b) by using finite element method (FEM) implemented in CST. As it is clear from this figure, although the field distribution is not perfectly cylindrical due to the discretization, the actual line source inside the cylindrical metasurface is cloaked by the metasurface and a fictitious line source is inspected by an observer outside the metasurface at electromagnetic large distances.

\section{Discussion and conclusions}
We have investigated the synthesis problem of cylindrical metasurfaces for general illusion mechanisms, and have presented a general analysis and synthesis approach by modeling metasurfaces with effective polarizability tensors. We have presented three practical examples including scattering cancellation from PEC and dielectric cylinders as cloaking examples. Furthermore we have also shown how to generate an illusion, a translated fictitious line source, from an original line source by covering such a line source with a cylindrical metasurface.\\
Besides the mentioned applications, the proposed approach is general and can be applied to many other problems. That is, any manipulation of electromagnetic waves including but not restricted to focusing lenses, designing antennas with arbitrary radiation patterns, beam forming, etc. Our study  opens up possibilities for complex electromagnetic wave manipulations using cylindrical topologies including  stealth technology, antenna synthesis problems, wireless power transfer systems, absorbers, etc.

\section*{Acknowledgment}
The authors would like to thank DS SIMULIA for providing CST Studio Suite that was instrumental in this study, and  Prof. C. Simovski and Prof. S. Tretyakov, from Aalto University, Finland,  for fruitful discussions.

\appendix
\section{Boundary conditions}\label{conditions}
Let us consider the proposed cylindrical metasurface of radius $\rho =a$ whose axis coincides with the $z$-axis (see Fig.~\ref{geom}). We start from the Maxwell's curl equations with  electric and magnetic polarization densities $\mathbf{P}_{\rm v}=\mathbf{P}\delta (\rho -a)$ and $\mathbf{M}_{\rm v}=\mathbf{M}\delta (\rho -a)$ located at $\rho=a$, i.e.,
\e  \nabla \times \mathbf{E}=-j\omega \left({{\mu_0 }}\mathbf{H}+\mathbf{M}_{\rm v}\right),\l{Max1} \f

\e  \nabla \times \mathbf{H}=j\omega \left(\epsilon_0 \mathbf{E}+\mathbf{P}_{\rm v}\right),\l{Max2} \f
\\
In the Eqs.~\r{Max1} and ~\r{Max2}, ${{\mu_0 }}$  and ${{\epsilon_0 }}$ are the free space magnetic permeability and electric permittivity, respectively, and we consider the constitutive relations $\_D=\epsilon_0\_E+\_P_{\rm v}$ and $\_B=\mu_0\_H+\_M_{\rm v}$ between the electric displacement $\_D$, magnetic induction $\_B$, electric field $\_E$, magnetic field $\_H$, electric polarization density $\mathbf{P}_{\rm v}$, and magnetic polarization density $\mathbf{M}_{\rm v}$. Next, since the problem has cylindrical symmetry, it is convenient to represent all vectors in cylindrical coordinates as
\e \mathbf{E}={{E}_{\rho}}\hat{\boldsymbol{\rho}}+{{E}_{\phi }}\hat{\boldsymbol{\phi} }+{{E}_{z}}\hat{\_z},\l{EC} \f 
\e \mathbf{H}={{H}_{\rho}}\hat{\boldsymbol{\rho}}+{{H}_{\phi }}\hat{\boldsymbol{\phi} }+{{H}_{z}}\hat{\_z}, \l{HC} \f
\e \mathbf{P}_{\rm v}={{P}_{{\rm v}\rho}}\hat{\boldsymbol{\rho}}+{{P}_{{\rm v}\phi }}\hat{\boldsymbol{\phi}}+{{P}_{{\rm v}z}}\hat{\_z},\l{PC} \f
\e \mathbf{M}_{\rm v}={{M}_{{\rm v}\rho}}\hat{\boldsymbol{\rho}}+{{M}_{{\rm v}\phi }}\hat{\boldsymbol{\phi}}+{{M}_{{\rm v}z}}\hat{\_z}.\l{MC}\f
\\
By using~\r{EC}--\r{MC} in the Maxwell curl equations~\r{Max1} and~\r{Max2} in cylindrical coordinate system we have
\e \frac{1}{\rho }\frac{\partial {{E}_{z}}}{\partial \phi }-\frac{\partial {{E}_{\phi }}}{\partial z}=-j\omega \left({{\mu_0 }}{H_\rho}+{M_{{\rm v}\rho}}\right),\l{curl1} \f

\e \frac{\partial {{E}_{\rho }}}{\partial z}-\frac{\partial {{E}_{z}}}{\partial \rho }=-j\omega \left({{\mu_0 }}{H_\phi}+{M_{{\rm v}\phi}}\right),\l{curl2} \f

\e \frac{1}{\rho }\frac{\partial \left(\rho {{E}_{\phi }}\right)}{\partial \rho }-\frac{1}{\rho }\frac{\partial {{E}_{\rho }}}{\partial \phi }=-j\omega \left({{\mu_0 }}{H_z}+{M_{{\rm v} z}}\right),\l{curl3} \f

\e \frac{1}{\rho }\frac{\partial {{H}_{z}}}{\partial \phi }-\frac{\partial {{H}_{\phi }}}{\partial z}=j\omega \left(\epsilon_0 {E_\rho}+{P_{{\rm v}\rho}}\right),\l{curl4} \f

\e \frac{\partial {{H}_{\rho }}}{\partial z}-\frac{\partial {{H}_{z}}}{\partial \rho }=j\omega \left(\epsilon_0 {E_\phi}+{P_{{\rm v}\phi}}\right),\l{curl5} \f

\e \frac{1}{\rho }\frac{\partial \left(\rho {{H}_{\phi }}\right)}{\partial \rho }-\frac{1}{\rho }\frac{\partial {{H}_{\rho }}}{\partial \phi }=j\omega \left(\epsilon_0 {E_z}+{P_{{\rm v}z}}\right).\l{curl6} \f
 
Then, we express the normal (with respect to the metasurface boundary) components of the electric and magnetic fields in terms of their tangential counterparts from~\r{curl4} and~\r{curl1}, respectively, as
\e {{E}_{\rho }}=\frac{1}{j\omega \epsilon_0}\left(\frac{1}{\rho }\frac{\partial {{H}_{z}}}{\partial \phi }-\frac{\partial {{H}_{\phi }}}{\partial z}\right)-\frac{{{P}_{{\rm v}\rho }}}{\epsilon_0 },\l{NTE} \f

\e {{H}_{\rho }}=-\frac{1}{j\omega \mu_0}\left(\frac{1}{\rho }\frac{\partial {{E}_{z}}}{\partial \phi }-\frac{\partial {{E}_{\phi }}}{\partial z}\right)-\frac{{{M}_{{\rm v}\rho }}}{\mu_0 }.\l{NTH} \f
One can obtain the boundary conditions for cylindrical metasurfaces given in~\r{BC1}--\r{BC4}, by substituting~\r{NTE} into Eqs.~\r{curl2} and~\r{curl3}, and~\r{NTH} into Eqs.~\r{curl5} and ~\r{curl6}, and then integrating the resulting equations over the metasurface thickness along the $\rho$-direction.


\begin{thebibliography}{11}

\bibitem{Kuester}
E. F. Kuester, M. Mohamed, M. Piket-May, and C. Holloway, IEEE Trans. Antennas Propag. {\bf 51}, 2641 (2003).

\bibitem{Holloway1}
C. Holloway, A. Dienstfrey, E. F. Kuester, J. F. O’Hara, A. K. Azad, and A. J. Taylor, Metamaterials {\bf 3}, 100 (2009).

\bibitem{Holloway2}
C. Holloway, E. F. Kuester, J. Gordon, J. O’Hara, J. Booth, and D. Smith, IEEE Antennas Propag. Mag. {\bf 54}, 10 (2012).


\bibitem{explain2}
The term ``metasurface'' is referred to surfaces with negligible electrical (optical)~\cite{def1} thickness~\cite{Holloway2}. Therefore, cylindrical surfaces as considered here (with negligible thickness in the radial direction) are considered in the category of metasurafces although it is a three-dimensional object.


\bibitem{kazemi2019simultaneous}
H. Kazemi, M. Albooyeh and F. Capolino, arXiv:1905.04439 (2019).

\bibitem{Padilla}
W. J. Padilla, M. T. Aronsson, C. Highstrete, M. Lee, A. J. Taylor, and R. D. Averitt, Phys. Rev. B {\bf 75}, 041102(R) (2007).

\bibitem{Caiazzo}
M. Caiazzo, S. Maci, and N. Engheta, IEEE Antennas Wirel. Propag. Lett. {\bf 3}, 261 (2004).

\bibitem{Chan}
 W. L. Chan, H.-T. Chen, A. J. Taylor, I. Brener, M. J. Cich, and D. M. Mittleman, Appl. Phys. Lett. {\bf 94}, 213511 (2009).

\bibitem{Pfeiffer1}
C. Pfeiffer and A. Grbic, Phys. Rev. Applied {\bf 2}, 044011 (2014).

\bibitem{Selv}
M. Selvanayagam and G. Eleftheriades, IEEE Trans. Antennas Propag. {\bf v}, 6155 (2014).

\bibitem{Capolino}
F. Capolino, A. Vallecchi, and M. Albani, IEEE Trans. Antennas Propag. {\bf 61}, 852 (2013).

\bibitem{Pfeiffer2}
C. Pfeiffer, C. Zhang, V. Ray, L.J. Guo,  and A. Grbic, Phys. Rev. Lett. {\bf 113}, 023902 (2014).

\bibitem{Zhao}
Y. Zhao, N. Engheta, and A. Al{\'u}, Metamaterials {\bf 5}, 90 (2011).

\bibitem{Albooyeh1}
M. Albooyeh, R. Alaee, C. Rockstuhl, and C. Simovski, Phys. Rev. B {\bf 91}, 195304 (2015).

\bibitem{Albooyeh2018Applications}
M. Albooyeh, V. Asadchy, J. Zeng, H. Kazemi, and F. Capolino,  arXiv:1811.04176 (2018).

\bibitem{Albooyeh2}
M. Albooyeh, Y. Ra'di, M.Q.  Adil, and C.R. Simovski, Phys. Rev. B {\bf 88}, 085435 (2013).

\bibitem{Asadchy1}
V.S. Asadchy, A. D{\'i}az-Rubio, S.N. Tcvetkova, D.H. Kwon, A. Elsakka, M. Albooyeh, and S. A. Tretyakov, Phys. Rev. X {\bf 7}, 031046 (2017).

\bibitem{Asadchy}
V.S. Asadchy, M. Albooyeh, S.N. Tcvetkova, A. D{\'i}az-Rubio, Y. Ra'di, and S.A. Tretyakov, Phys. Rev. B {\bf 94}, 075142 (2016).

\bibitem{Epstein1}
A. Epstein and G. V. Eleftheriades, J. Opt. Soc. Am. B {\bf 33}, A31-A50 (2016).

\bibitem{Epstein2}
J. PS Wong, A. Epstein, and G. V. Eleftheriades, IEEE Antennas Wirel. Propag. Lett. {\bf 15}, 1293 (2016).


\bibitem{Dana}
M. Danaeifar, N. Granpayeh, N. A. Mortensen, and S. Xiao, J. Phys. D {\bf 48}, 385106 (2015).

\bibitem{Niemi}
T. Niemi, A. Karilainen, and S. Tretyakov, IEEE Trans. Antennas Propag. {\bf 61}, 3102 (2013).

\bibitem{Salem}
M. A. Salem and C. Caloz, Opt. Express {\bf 22}, 14 530 (2014).

\bibitem{Kazemi2019Perfect}
H. Kazemi, M. Albooyeh, and F. Capolino, in URSI EM Theory Symposium, EMTS 2019, San Diego, CA, 27–31
May 2019.

\bibitem{Achouri}
K. Achouri, M. A. Salem, and C. Caloz, IEEE Trans.  Antennas  Propag. {\bf 63}, 2977 (2015).

\bibitem{Albooyeh3}
M. Albooyeh, D. Morits, and C. R. Simovski, Metamaterials {\bf 5}, 178 (2011).


\bibitem{Albooyeh31}
M. Albooyeh, S. Tretyakov, and C. Simovski, Ann. der Phys. {\bf 528}, 721 (2016).

\bibitem{Roberts}
C.M. Roberts, S. Inampudi, and V.A. Podolskiy, Opt. Express {\bf 23}, 2764 (2015).
\bibitem{Kiani}
M. Kiani, A. Abdolali, and M.M. Salary, IEEE Trans. Antennas Propag. {\bf 63}, 631118 (2015).

\bibitem{Oraizi}
H. Oraizi and A. Abdolali, Prog. Electromagn. Res. B {\bf 3}, 227 (2008).

\bibitem{Alaeeabs}
R. Alaee, C. Menzel, C. Rockstuhl, and F. Lederer, Opt. Express {\bf 20}, 18370 (2012).


\bibitem{Kishk}
A. A. Kishk and P-S. Kildal, IEEE Trans. Antennas Propag. {\bf 45}, 51 (1997).

\bibitem{Raffaelli}
S. Raffaelli, Z. Sipus, and P-S. Kildal, IEEE trans Antennas Propag. {\bf 53}, 1105 (2005).


\bibitem{Padooru}
Y.R. Padooru, A.B. Yakovlev, Pai-Yen Chen, and Andrea Al{\'u}, J. Appl. Phys. {\bf 112}, 034907 (2012).

\bibitem{Padooru1}
Y.R. Padooru, A.B. Yakovlev, Pai-Yen Chen, and A. Al{\'u}, J. Appl. Phys. {\bf 112}, 104902 (2012).


\bibitem{Bernety}
H.M. Bernety, and A.B. Yakovlev, IEEE Trans. Antennas Propag. {\bf 63}, 1554 (2015).

\bibitem{Raeker}
B. O. Raeker and S. M. Rudolph, IEEE Antennas Wirel. Propag. Lett. {\bf 15}, 1101 (2016).


\bibitem{Raeker2}
B. O. Raeker and S. M. Rudolph, IEEE Antennas Wirel. Propag. Lett. {\bf 16}, 995 (2017).

\bibitem{jia}
J. Xiao, Y. Vahabzadeh, C. Caloz, Y. Fan, IEEE Trans. Antennas Propag. 67, 2542 (2019).

\bibitem{Dehmol}
M. Dehmollaian, N. Chamanara, C. Caloz, IEEE Trans. Antennas Propag. doi: 10.1109/TAP.2019.2905711 (2019).


\bibitem{Tretyakov}
S.A. Tretyakov {\it Analytical Modeling in Applied Electromagnetics} (Artech House, Norwood, MA, 2003). 


\bibitem{Serdyukov}
A.~Serdyukov,	I.~Semchenko,	S.~Tretyakov,	and A.~Sihvola, {\it Electromagnetics of Bi-anisotropic Materials: Theory and Applications} (Gordon and Breach Science Publishers, Amsterdam, 2001).

\bibitem{Idemen_book}
M. Idemen, {\it Universal boundary conditions and Cauchy data for the electromagnetic field} in {\it Essays on the formal aspects of electromagnetic theory} (World Scientific, New Jersey, 1993), pp. 657--698.


\bibitem{felsen}
L. B. Felsen and N. Marcuvitz, {\it Radiation and Scattering of Waves} (Wiley, New York, 1994), pp. 185--187.


\bibitem{AlbooyehEqui}
M. Albooyeh, D.H. Kwon, F. Capolino, and S.A. Tretyakov, Phys. Rev. B {\bf 95}, 115435 (2017).

\bibitem{Albooyeh_Nromal}
M. Albooyeh, H. Kazemi, F. Capolino, D.H. Kwon, and S.A. Tretyakov, in 2017 IEEE International Symposium on Antennas and Propagation \& USNC/URSI National Radio Science Meeting, pp. 1707-1708 (2017).

\bibitem{disc1}
Note that although there is no theoretical limitation to realize any desired electromagnetic field with only tangetial (to the surface) polarization densities, there would be extra {\it practical} degrees of freedom in realization of a desired field using normal polarization densities as discussed in Refs.~\cite{AlbooyehEqui,pfeiffer2016emulating}.


\bibitem{balanis}
C.A. Balanis, {\it Advanced Engineering Electromagnetics} (John Wiley \& Sons, Danvers, MA, 2012).

  
\bibitem{Chen}
P. Y. Chen and A. Al{\'u}, Phys. Rev. B, {\bf 84}, 205110 (2011).

\bibitem{Schurig}
D. Schurig, J.J. Mock, B.J. Justice, S.A. Cummer, J.B. Pendry, A.F. Starr, and D.R. Smith, Science {\bf 314}, 977 (2006).

\bibitem{Yi}
J. Yi, P.H. Tichit, S.N. Burokur, and A. de Lustrac, J. Appl. Phys. {\bf 117}, 084903 (2006).


\bibitem{Tretyakov1}
S. Tretyakov, P. Alitalo, O. Luukkonen, and C. Simovski, Phys. Rev. Lett. {\bf 103}, 103905 (2009). 


\bibitem{Kwon}
D.H. Kwon and D. M. Pozar, IEEE Trans. Antennas Propag. {\bf 57}, 3720 (2009).


\bibitem{albooyeh_abs}
M. Albooyeh and C. R. Simovski, Opt. Express {\bf 20}, 21888 (2012).

\bibitem{radi_abs1}
Y. Ra'di, V.S. Asadchy, and S.A. Tretyakov, IEEE Trans. Antennas Propag {\bf 61}, 4606 (2013).

\bibitem{radi_abs2}
Y. Ra'di, C.R. Simovski, and S.A. Tretyakov, Phys. Rev. Appl. {\bf 3}, 037001 (2015).

\bibitem{alaee_abs}
R. Alaee, M. Albooyeh, and C. Rockstuhl, J. Phys. D: Appl. Phys. {\bf 50}, 503002 (2017).

\bibitem{COMSOL}
\url{https://www.comsol.com/}

\bibitem{Harrington}
R. F. Harrington, {\it Time-Harmonic Electromagnetic Fields} (McGraw-Hill, New York, 1961).

\bibitem{Song}
M. Song, P. Belov, and P. Kapitanova, Appl. Phys. Rev. {\bf 4}, 021102 (2017).

\bibitem{def1}
The term ''electrically (optically) thin'' refers to the thickness as compared to the wavelength.


\bibitem{safari}
M. Safari, A. Abdolali, H. Kazemi, M. Albooyeh, M. Veysi, and F. Capolino, in 2017 IEEE International Symposium on Antennas and Propagation \& USNC/URSI National Radio Science Meeting, pp. 1499–1500 (2017).


\bibitem{pfeiffer2016emulating}
C. Pfeiffer and A. Grbic, Phys. Rev. Lett. {\bf 117}, {077401} (2016).
  
\end{thebibliography}
\end{document}